\documentclass[10pt, conference, letterpaper]{IEEEtran}
\IEEEoverridecommandlockouts
\usepackage{cite}
\usepackage{amsmath,amssymb,amsfonts}
\usepackage{algorithm}
\usepackage{algorithmic}
\usepackage{graphicx}
\usepackage{textcomp}
\usepackage{xcolor}
\usepackage[caption=false,font=footnotesize]{subfig}
\def\BibTeX{{\rm B\kern-.05em{\sc i\kern-.025em b}\kern-.08em
    T\kern-.1667em\lower.7ex\hbox{E}\kern-.125emX}}
\begin{document}

\title{THz-SynC: Collective Synthesis with Contextual-Bandit-Assisted Coordination for Reconfigurable Hybrid Optical–THz AI Datacenters
}

\author{Jingting Jiang and Chong Han
    \\Shanghai Jiao Tong University, Shanghai, China.
    \\E-mail: {jingting.jiang, chong.han}@sjtu.edu.cn}
\maketitle

\begin{abstract}
Terahertz (THz) wireless interconnects offer high-capacity, low-latency, and energy-efficient rack-to-rack links capable of on-demand connectivity reconfiguration, serving as a promising complement to optical fabrics for communication-intensive distributed AI training datacenters.
However, co-optimizing optical and THz resources to minimize collective completion time and transmission energy remains challenging due to dynamic optical congestion, THz channel fluctuations, and heterogeneous compute stragglers.
Existing reconfigurable data-center designs predominantly optimize network topology and traffic routing, with limited consideration of collective communication semantics in distributed AI workloads over hybrid fabrics.
To address these challenges, we propose THz-SynC, a novel framework that integrates collective synthesis with contextual-bandit-assisted hybrid-fabric coordination to optimize the tradeoff between collective completion time and transmission energy.
By exploiting collective-specific semantics, THz-SynC synthesizes tailored communication topologies for All-to-All and AllReduce patterns while dynamically allocating THz resources.
Furthermore, a contextual-bandit coordinator adaptively routes communication chunks across optical and THz links and selects rack power budgets leveraging real-time observations of network states and collective semantics.
Trace-driven evaluations show that THz-SynC outperforms wired-only, wireless-only, and hybrid baselines, achieving a superior delay–energy Pareto frontier under dynamic network conditions.

% Terahertz (THz) wireless interconnects provide direct, high-capacity, and energy-efficient rack-to-rack links that can reconfigure connectivity on demand, offering a promising complement to wired fabrics for communication-intensive distributed AI training in datacenter. 
% However, jointly scheduling wired and THz resources to reduce collective completion time and transmission energy is difficult under time-varying wired congestion, wireless channel fluctuations, and compute stragglers. 
% Existing reconfigurable data-center designs largely optimize network topology and traffic routing, with limited consideration of the collective communication semantics of distributed AI training, particularly in hybrid wired–THz fabrics.
% We present THzColl, a framework that jointly performs wireless-aware collective synthesis and hybrid-fabric coordination to optimize the latency and transmission energy tradeoff of distributed AI training. 
% THzColl develops collective-specific synthesis mechanisms for All-to-All and AllReduce, together with resource allocation tailored to the THz wireless links.
% It further uses a contextual-bandit-assisted coordinator to assign communication chunks to the wired or THz links based on online observations of network environments and collective semantics. 
% We evaluate THzColl using distributed training traces spanning different models under varying network fluctuations. Results show that THzColl achieves a more favorable collective completion time and energy tradeoff than wired-only, wireless-only, and hybrid baselines.
\end{abstract}

% THzColl synthesizes All-to-All communication as a sequence of directed matchings and realizes AllReduce through multiple rooted, fanout-bounded trees, together with THz resource allocation. 


\section{Introduction}

Artificial intelligence models are increasingly trained across large clusters of GPUs and specialized accelerators. As model size, expert count, and parallelism degree continue to grow, training performance depends not only on computation throughput but also on the ability of the interconnect to move activations, tokens, and gradients at scale \cite{jiang2024megascale, shi2021towards}. 
These transfers are dominated by collective communication. Pipeline parallelism (PP) generates point-to-point (P2P) activation and gradient exchanges, mixture-of-experts (MoE) models introduce irregular All-to-All (A2A) traffic, and data parallelism requires repeated AllReduce (AR) synchronization. The completion time of these collectives directly affects pipeline stalls and iteration time, which makes the network an important design component of distributed AI systems.

Current AI clusters primarily rely on electrical and optical fabrics. These technologies provide high capacity and mature reliability, but their fixed cabling and switch hierarchy limit how rapidly connectivity can follow changing communication demands. Bursty collective traffic can concentrate load on shared links and buffers, producing queueing and backpressure even when the nominal link rate is high. Wired interconnects incur substantial cabling complexity and deployment cost, and become increasingly difficult to scale as AI clusters expand. Reconfigurable optical networks improve adaptability by adjusting logical connectivity, but remain constrained by circuit configuration overhead and multi-hop forwarding. These limitations motivate a more flexible fabric that can establish direct rack-to-rack links according to the communication demand.

Wireless interconnects provide such reconfigurability without rewiring. Millimeter-wave and optical wireless technologies have therefore been investigated for data center networks \cite{halperin2011augmenting, hamedazimi2014firefly}. 
Among emerging wireless technologies, terahertz (THz) communication \cite{akyildiz2022terahertz} is particularly well suited to wireless data centers, especially for communication-intensive AI training workloads \cite{han2026wires}. The wide spectrum can support substantially higher peak rates than millimeter-wave systems, which is increasingly important as rapidly growing model sizes generate massive gradient-communication demands during distributed training.
Compared with optical wireless links, which are more susceptible to dust and partial obstructions \cite{ghobadi2016projector}, THz links can provide more robust connectivity. 
Furthermore, in suitable short-range applications, THz links also have the potential to reduce energy per bit by bypassing multiple switching stages and avoiding repeated electro-optical conversion \cite{han2026wires}.

However, exploiting a THz overlay for distributed training remains an open problem. 
Existing reconfigurable or wireless data center designs mainly optimize network topology and traffic routing based on aggregate demand matrices \cite{mao2026atro, ghobadi2016projector}. Such abstractions aren’t well suited to distributed training collectives, whose execution depends not only on path selection but also on specific communication semantics. 
In parallel, collective communication systems \cite{shah2023taccl} optimize schedules for primitives such as AR and A2A, but most assume fixed electrical or optical fabrics and overlook how on-demand wireless links can complement the wired network. Consequently, collective-aware communication synthesis for hybrid wired–wireless coordination remains unexplored.
Furthermore, the time-varying wired congestion, THz channel fluctuations, and compute stragglers have increased the challenges of coordinating optical and THz fabrics. 

To address these problems, we present THz-SynC, a collective-synthesis and hybrid-fabric coordination framework for distributed AI training over reconfigurable optical and THz data-center networks.
THz-SynC first synthesizes collective-specific communication structures that account for the characteristics of THz networks. For each synthesized flow, a contextual-bandit coordinator determines the number of chunks assigned to the optical and THz fabrics, together with the transmission power budget of each participating rack based on the current network conditions. More specifically, wired chunks traverse the optical network, whereas wireless chunks are served through multi-round scheduling and wireless resource allocation. Under the selected rack-level power budgets, the wireless scheduler minimizes round completion time while accounting for heterogeneous compute stragglers.

The main contributions are summarized as follows

\begin{itemize}
    \item We develop collective-aware synthesis for a reconfigurable THz overlay. The synthesized structures account for heterogeneous compute stragglers and enable concurrent wireless transmissions without violating collective semantics.
    \item We design a contextual-bandit-assisted coordinator for hybrid optical and THz fabrics. It jointly determines per-flow wired and wireless chunk ownership and rack-level THz power budgets from dynamic network and workload context. 
    \item We build a trace-driven simulation that integrates Megatron-LM-style \cite{shoeybi2019megatron} training dependencies. Simulation results demonstrate that THz-SynC achieves a better delay–energy tradeoff than the baselines.
\end{itemize}

\section{Related Work}

\textbf{Reconfigurable Optical Datacenters.} Fiber-based reconfigurable datacenter networks use optical circuit switches (OCSs) to reshape rack-level connectivity as traffic demand changes. c-Through and Helios augment packet-switched fabrics with optical circuits, placing high-volume transfers on direct optical paths while retaining an electrical network for general traffic \cite{wang2010c, farrington2010helios}. Instead, RotorNet \cite{mellette2017rotornet} avoids demand-driven circuit assignment by cycling through a predetermined sequence of matchings, whereas Opera \cite{mellette2019expanding} maintains a time-varying expander and lets bulk traffic wait for direct circuits while forwarding latency-sensitive traffic over available multi-hop paths. Sirius \cite{ballani2020sirius} further co-designs a flat optical fabric with nanosecond-scale switching, routing, scheduling, congestion control, and synchronization. These systems span traffic-aware hybrid fabrics, periodic reconfiguration, and fast all-optical switching, but their control interfaces generally operate on demand matrices or circuit schedules rather than the dependency and completion semantics of training collectives. 

\textbf{Wireless Datacenters.} Existing wireless data-center designs primarily rely on millimeter-wave and free-space optical (FSO) technologies, using directional links either as temporary shortcuts over a wired fabric or as the primary rack-to-rack interconnect. As complements to wired networks, \cite{halperin2011augmenting} introduces multi-gigabit 60-GHz flyways to augment capacity between congested rack pairs, while \cite{zhang20113d} proposes a 3D beamforming architecture that reflects directional signals from the ceiling to extend communication range and mitigate blockage and interference. A fully wireless design in \cite{shin2012feasibility} instead arranges 60-GHz transceivers and racks around a Cayley-graph topology. FireFly \cite{hamedazimi2014firefly} and ProjecToR \cite{ghobadi2016projector} replace radio-frequency links with steerable FSO links to create reconfigurable rack-to-rack connectivity.

\textbf{Collective Communication Optimization and Synthesis.} Collective communication research tailors communication algorithms and training fabrics to the topology of wired GPU clusters. For instance, SCCL \cite{cai2021synthesizing} encodes topology-specific collective synthesis as an SMT problem and searches for latency- and bandwidth-efficient algorithms. Blink \cite{wang2019blink} constructs collective primitives by packing spanning trees over heterogeneous links. TACCL \cite{shah2023taccl} introduces communication sketches that constrain routing and ordering choices, enabling synthesis across multi-node heterogeneous topologies. Beyond collective libraries, TopoOpt \cite{wang2023topoopt} co-optimizes DNN parallelization, routing, and a reconfigurable wired optical topology. 
Additionally, Meta's RoCE deployment \cite{gangidi2024rdma} adapts topology, routing, transport, and collective-library behavior to production training traffic. 
Although existing studies optimize collective communication through topology, routing, and scheduling, integrating collectives with more flexible wireless connectivity remains insufficiently explored.

\section{System Model and Problem Formulation}
\label{sec:system}
 We consider a distributed AI training cluster interconnected by a wired optical fabric and a reconfigurable THz wireless overlay. The optical fabric provides persistent baseline connectivity, while network reconfiguration is realized primarily through the THz overlay, which establishes directional single-hop rack-to-rack links on demand.

\begin{figure}[!t]
    \centering
    \includegraphics[width=0.8\columnwidth]{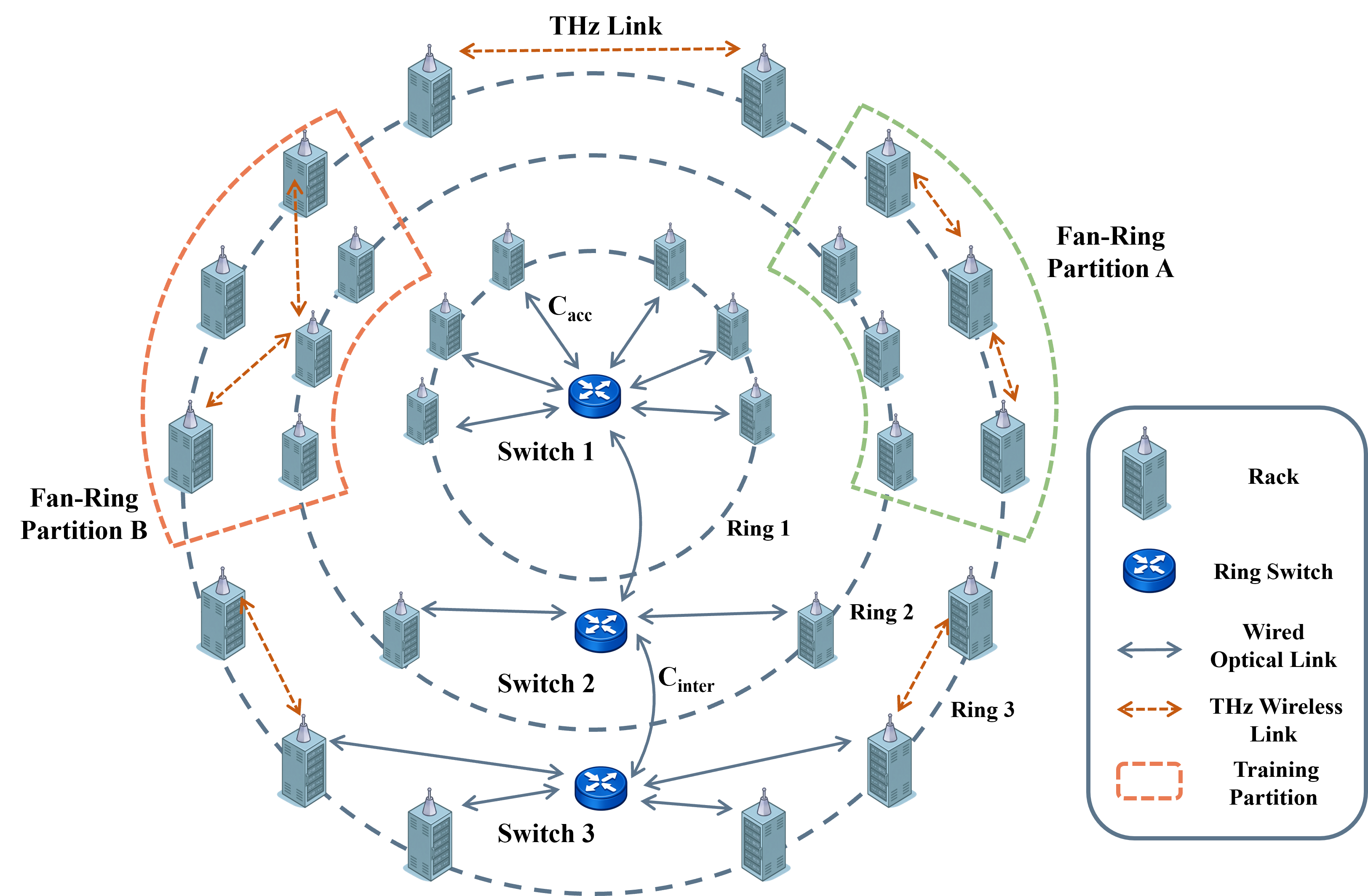}
    \caption{Hybrid Optical–THz Data center.}
    \label{fig:system model}
\end{figure}

\subsection{Hybrid Optical–THz Data centers}

As illustrated in Fig.~\ref{fig:system model}, the data center arranges racks on multiple concentric rings. We focus on the datacenter containing \(S\) rings, indexed by \(s\in\{1,\ldots,S\}\), and \(R\) racks per ring, indexed by \(r\in\{1,\ldots,R\}\). Rack \(v_{s,r}\) has global index \(i=sR+r\). The radius of ring \(s\) is
\(\rho_s=\rho_0+s\Delta\rho\), where \(\rho_0\) is the innermost radius and \(\Delta\rho\) is the radial spacing between adjacent rings. 
All rings share the same angular grid, so racks with the same index \(r\) across different rings lie along the same radial line. Let \(\Delta\theta\) denote the fixed angular spacing between consecutive racks. Therefore, given two racks \(i=v_{s,r}\) and \(j=v_{s',r'}\), the distance can be calculated by
\begin{equation}
d_{ij}
=
\sqrt{
\rho_s^2+\rho_{s'}^2
-2\rho_s\rho_{s'}
\cos\!\left((r-r')\Delta\theta\right)
},
\label{eq:rack_distance}
\end{equation}
Since all rack-top transceivers are installed at the same height, the vertical component does not affect \(d_{ij}\).
This ring-shaped placement reduces the rack-to-rack blockage commonly observed in regular row-column layouts. More importantly, racks distributed along the same ring can maintain line-of-sight (LoS) paths in the circumferential direction, enabling efficient direct inter-rack THz communication. This property is particularly aligned with distributed learning workloads, whose collective operations frequently involve concurrent data exchanges among multiple racks.

The data center may concurrently host multiple training jobs. We suppose that the data center is divided into multiple fan-ring partitions to support different training jobs. Each partition contains a subset of rings and a selected set of racks on those rings. Let \(\mathcal{V}_m\) denote the racks assigned to job \(m\). The provisioning layer allocates disjoint rack and communication resources to different jobs, such that $\mathcal{V}_m \cap  \mathcal{V}_{m'}=\varnothing,\qquad m\neq m'$.
We therefore focus on one representative partition, and the THz-SynC procedure can be executed independently in the remaining partitions without resource conflicts.
Within a partition, PP stages are mapped to distinct rings. Racks with the same lane index \(r\) across these rings host the corresponding PP ranks and form a radial PP lane. Racks on ring \(s\) host different DP ranks of PP stage \(s\). Hence, forward activations and backward gradients are primarily exchanged across adjacent rings, whereas DP collectives, including MoE A2A and gradient AR, primarily operate within a ring.

The optical fabric provides persistent baseline connectivity and operates in full-duplex mode. Each ring \(s\) is associated with a switch \(w_s\). Every rack on ring \(s\) connects to \(w_s\) through an optical fiber of capacity \(C_{\mathrm{acc}}\), while the ring switches are interconnected by optical fibers of capacity \(C_{\mathrm{inter}}\). For racks \(i=v_{s,r}\) and \(j=v_{s',r'}\), the wired path is
\begin{equation}
\mathcal{P}_{ij}^{W} =
\begin{cases}
(i,w_s,j), & s=s',\\
(i,w_s,w_{s'},j), & s\neq s'.
\end{cases}
\label{eq:wired_path}
\end{equation}
Each rack-facing port maintains a queue of capacity \(B_{\mathrm{nic}}\), and each switch output maintains a queue of capacity \(B_{\mathrm{out}}\). Switch \(w_s\) also provides a shared buffer of capacity \(B_{s}\) for its output queues. Let \(Q_{\ell}(t)\) denote the occupancy of the queue associated with directed link \(\ell\). The finite-buffer constraints are
\begin{equation}
0 \leq Q_{\ell}(t) \leq B_{\ell},
\qquad
\sum_{\ell\in\delta^{+}(w_s)} Q_{\ell}(t)
\leq B_{s},
\label{eq:wired_buffers}
\end{equation}
where \(B_{\ell}=B_{\mathrm{nic}}\) for rack-facing queues and
\(B_{\ell}=B_{\mathrm{out}}\) for switch output queues. The set
\(\delta^{+}(w_s)\) contains the directed output links of \(w_s\). Consider a wired flow \(f\) released at time \(t_f^{\mathrm{rel}}\), transmitted from rack \(i\) to rack \(j\), and divided into a packet set \(\mathcal{K}_f\). Its completion time is determined by the arrival of its last packet. Therefore, the actual wired transmission delay is given by
\begin{equation}
T_f^{O}
=
\max_{k\in\mathcal{K}_f}
\left[
T_{f,k}^{\mathrm{adm}}
+
\sum_{\ell\in\mathcal{P}_{ij}^{W}}
\left(
T_{f,k,\ell}^{\mathrm{q}}
+
\frac{L_{f,k}}{C_{\ell}}
+
\tau_{\ell}^{\mathrm{sw}}
\right)
\right],
\label{eq:wired_flow_delay}
\end{equation}
where \(L_{f,k}\) is the size of packet \(k\), \(C_{\ell}\) is the capacity of link \(\ell\), and \(T_{f,k,\ell}^{\mathrm{q}}\) is the queueing delay caused by other traffic. The term \(\tau_{\ell}^{\mathrm{sw}}\) denotes switch-processing delays. The admission delay \(T_{f,k}^{\mathrm{adm}}\) captures the waiting time incurred when the corresponding port queue or shared buffer lacks sufficient space. This formulation follows store-and-forward forwarding and neglects propagation delay in the optical fiber.

Furthermore, each rack carries a THz transceiver at its top. 
The THz wireless network can provide on-demand one-hop connectivity that bypasses multihop wired forwarding, and offloads congested wired paths according to collective traffic demands.
Based on the channel measurements reported in \cite{han2026wires}, the LoS and non-line-of-sight (NLoS) path losses between racks \(i\) and \(j\) are modeled as

\begin{equation}
\begin{aligned}
PL_{\mathrm{LoS}}(d_{ij},f)
&=
18.8\log_{10}d_{ij}
+82.69
+20\log_{10}\left(\frac{f}{f_0}\right),\\
PL_{\mathrm{NLoS}}(d_{ij},f)
&=
7.6\log_{10}d_{ij}
+106.6
+20\log_{10}\left(\frac{f}{f_0}\right).
\end{aligned}
\label{eq:measured_path_loss}
\end{equation}
Here, \(f_0\) is the reference frequency. The model is applicable to the \(290\)--\(310\) GHz frequency band.
The resulting channel power gain is given by
\begin{equation}
g_{ij}
={}
10^{-PL_{\mathrm{LoS}}(d_{ij},f)/10} 
+
\sum_n
10^{-PL_{\mathrm{NLoS,n}}(d_{ij},f)/10}.
\label{eq:composite_channel_gain}
\end{equation}
The NLoS term represents the aggregate contribution of indirect propagation paths characterized by the measurements in \cite{han2026wires}. 
Each THz transceiver supports \(N_c\) non-overlapping subbands, denoted by
\(\mathcal{C}=\{1,\ldots,N_c\}\), where each subband \(c\) has same bandwidth \(B_c\). Let \(x_{ij}^{c}(t)\in\{0,1\}\) indicate whether rack \(i\) transmits to rack \(j\) on subband \(c\). Each rack operates in half-duplex mode on an individual subband, which follows
\begin{equation}
\sum_{j\neq i}
\left(
b_{ij}^{c}(t)+b_{ji}^{c}(t)
\right)
\leq 1,
\qquad
\forall i,\ c\in\mathcal{C}.
\label{eq:rf_half_duplex}
\end{equation}
Given the transmission power \(p_{ij}^{c}(t)\) allocated to link \(i\rightarrow j\) on subband \(c\), its achievable rate can be calculated as
\begin{equation}
r_{ij}^{c}(t)
=
B_c\log_2\!\left(
1+
\frac{
p_{ij}^{c}(t)Gg_{ij}^{c}(t)
}{
N_0B_c+I_{ij}^{c}(t)
}
\right),
\label{eq:thz_rate}
\end{equation}
where \(G\) is the receiving antenna gain and \(N_0\) denotes the noise power spectral density. Owing to the high directionality of THz beams, the term \(I_{ij}^{c}(t)\), which denotes residual inter-link interference, can be treated as additional noise \cite{zhang2019joint}. 

% The proposed architecture supports incremental scale-up and scale-down without modifying the existing network infrastructure. Since each ring contains a fixed number of racks and is served by a switch with a corresponding fixed port count, capacity can be adjusted by adding or removing complete rack-ring units together with their associated switches. Existing switches and optical connections remain unchanged. Each rack is equipped with one THz transceiver, so extending the wireless overlay requires only the installation or removal of rack-local transceivers and does not introduce additional inter-rack cabling complexity.

\subsection{Training Trace and Collective Communication}

% Large-scale model training distributes computation and model state across GPUs using complementary parallelism dimensions. In DP, multiple model replicas process different input samples and synchronize their gradients after backward propagation. PP partitions consecutive model layers into stages and divides each training iteration into microbatches, allowing different stages to execute concurrently. Adjacent PP stages exchange forward activations and backward gradients. We confine TP groups to a rack and therefore exclude their intra-rack communication from the inter-rack optimization considered in this work.

% The resulting training execution contains three main inter-rack communication classes. Point-to-Point (P2P) communication transfers activations in the forward direction and gradients in the reverse direction between adjacent PP stages. All-to-All (A2A) communication arises primarily from token dispatch and combine operations in Mixture-of-Experts (MoE) models. In an A2A operation, each participant may send a distinct payload to every other participant. AllReduce (AR) instead synchronizes DP gradients. Each participant contributes a local tensor to a reduction, and the reduced result is returned to every participant. 
Large-scale training usually combines DP, PP, and tensor parallelism (TP). We assume TP communication remains within each rack and is efficiently supported by mature NCCL implementations over high-bandwidth intra-rack fabrics. We therefore focus on the inter-rack bottleneck, including P2P exchanges between adjacent pipeline stages, A2A token dispatch and combine in MoE models, and AR gradient synchronization across DP replicas.
We represent the communication behavior of one or more training iterations by a training trace
\begin{equation}
\mathcal{T}=\left(\mathcal{E},\prec\right),\qquad
\mathcal{E}=\left\{e_1,e_2,\ldots,e_{|\mathcal{E}|}\right\},
\label{eq:training_trace}
\end{equation}
where \(\mathcal{E}\) is the set of inter-rack communication events and \(\prec\) is the strict partial order induced by the training computation graph. 
An edge \(e'\prec e\) indicates that event \(e\) cannot become executable before \(e'\) completes. A chain such as \(e_1\rightarrow e_2\rightarrow\cdots\) describes one dependency path, but the complete trace is a directed acyclic graph that permits branches and concurrent events. 
% For example, Megatron-LM-style pipeline execution naturally produces interleaved forward P2P, backward P2P, and gradient-synchronization events across microbatches. Model architecture and parallelism choices introduce further variation, as a dense model such as Bidirectional Encoder Representations from Transformers (BERT) does not generate MoE token exchanges, whereas an MoE model introduces A2A dispatch and combine events. Their order and payload sizes follow from the model structure, microbatch execution, and selected PP and DP degrees.
Each event can be described as
\begin{equation}
e=\left(\tau_e,\mathcal{V}_e,\mathcal{D}_e,
r_e,\operatorname{Pred}(e)\right),
\label{eq:communication_event}
\end{equation}
where \(\tau_e\in\{\mathrm{P2P},\mathrm{A2A},\mathrm{AR}\}\) is the communication type, \(\mathcal{V}_e\) is the participating rack set, and \(\mathcal{D}_e\) is its semantic demand descriptor. For P2P and A2A, \(\mathcal{D}_e\) specifies the directed byte demand \(D_{ij}^{e}\) from source rack \(i\) to destination rack \(j\). For AR, \(\mathcal{D}_e\) records the local tensor size and reduction semantics, while the resulting source-destination transfers depend on the adopted collective synthesis scheme.
The nominal release time \(r_e\) is obtained from the trace, and
\(\operatorname{Pred}(e)=\{e'\in\mathcal{E}:e'\prec e\}\) contains the preceding communication events. 
% Our optimization covers every event in \(\mathcal{E}\). 

\subsection{Problem Formulation}

In this work, our target is to jointly optimize the completion time and transmission energy of all communication events in the training trace. However, the optimization must account for the time-varying service conditions of both the optical and THz fabrics. We denote the network state at time \(t\) by \(\boldsymbol{\omega}(t)\). 
In the optical fabric, background traffic, including control signaling, network log, and traffic from other partitions sharing the same switch infrastructure, arrives according to a Poisson process and shares finite output queues and switch buffers. The resulting queue occupancies evolve with packet arrivals and link service, which produces time-varying admission and queueing delays. 
The THz wireless links may experience time-varying channel deterioration due to blockage and propagation fluctuations.
Another practical factor is the presence of compute stragglers \cite{lin2025understanding}, which delay the availability of data generated by individual racks. Let \(\delta_{i,e}^{\mathrm{str}}\geq 0\) denote the straggler delay of rack \(i\) during event \(e\), the release time of a flow generated at rack \(i_f\) is $r_{f,k} = r_e + \delta_{i_f,e}^{\mathrm{str}}$.
Given the time-varying conditions of both fabrics and their distinct data rate and energy characteristics, optical and THz resources must be jointly scheduled.

For event \(e\), let \(\mathcal{F}_e\) denote its logical flow set. Each flow \(f\in\mathcal{F}_e\) carries \(D_f\) data units from rack \(i_f\) to rack \(j_f\) and is partitioned into
\(K_f=\lceil D_f/B_{\mathrm{ch}}\rceil\) chunks, where each chunk carries at most \(B_{\mathrm{ch}}\) data units. 
The chunk set is denoted by
\(\mathcal{K}_f=\{1,\ldots,K_f\}\). For each chunk \(k\in\mathcal{K}_f\), we define two binary variables \(x_{f,k}^{O}\) and \(x_{f,k}^{T}\) that indicate its assignment to the optical and THz fabrics, respectively.
Traffic splitting is performed independently for each logical flow. Therefore, the completion time of flow \(f\) can be described as
\begin{equation}
T_f
=
\max
\left\{
T_f^{O},
T_f^{T}
\right\}.
\label{eq:hybrid_flow_completion}
\end{equation}
Let \(f^{O}\) and \(f^{T}\) denote the subflows composed of the chunks satisfying \(x_{f,k}^{O}=1\) and \(x_{f,k}^{T}=1\), respectively. The optical completion time \(T_f^{O}\) follows the delay model defined in \ref{eq:wired_flow_delay}. The THz completion time is expressed as
\begin{equation}
T_f^{T}
=
\Phi_f^{T}
\left(
f^{T},
\mathbf{s}_e,
\mathbf{h}_e,
\mathbf{b}_e,
\mathbf{p}_e,
\boldsymbol{\omega}(t)
\right),
\label{eq:thz_flow_completion}
\end{equation}
where \(\mathbf{s}_e\), \(\mathbf{h}_e\), \(\mathbf{b}_e\), and \(\mathbf{p}_e\) denote the collective synthesis structure, transmission schedule, subband allocation, and power allocation, respectively. The function also depends on the realized network state \(\boldsymbol{\omega}(t)\). The collective-specific formulation of \(\Phi_f^{T}(\cdot)\) is presented in the Sec. \ref{sec:thz_sync}.

The dynamic transmission energy of event \(e\) consists of optical and THz components, which is
\begin{equation}
E_e
=
E_e^{O}
+
E_e^{T}.
\label{eq:event_energy_total}
\end{equation}
The optical transmission energy can be calculated as
\begin{equation}
E_e^{O}
=
\epsilon^{O}
\sum_{f\in\mathcal{F}_e}
D_f^{O},
\label{eq:optical_energy}
\end{equation}
where \(\epsilon^{O}\) is the fixed optical transmission-energy coefficient per data unit ~\cite{} and $D_f^{O}$ denote the data volume of flow \(f\) assigned to the optical fabric.
Let \(\mathcal{H}_e\) denote the THz transmission stages of event \(e\), and \(\mathcal{A}_{e,h}\) contain the transmissions activated during stage \(h\). The THz transmission energy is
\begin{equation}
E_e^{T}
=
\sum_{h\in\mathcal{H}_e}
\Delta_{e,h}
\sum_{u\in\mathcal{A}_{e,h}}
p_{u,e,h},
\label{eq:thz_energy}
\end{equation}
where \(\Delta_{e,h}\) is the duration of stage \(h\) and \(p_{u,e,h}\) is the transmit power allocated to transmission \(u\). 

Based on the preceding completion-time and energy models, we formulate a trace-level optimization that jointly determines chunk assignment, collective execution, and THz resource allocation. We define the fabric-assignment variables as
\(\boldsymbol{x}=\{x_{f,k}^{O},x_{f,k}^{T}\}\) and collect the transmission scheduling, subband allocation, and power allocation decisions in
\(\boldsymbol{a}=\{\mathbf{h}_e,\mathbf{b}_e,\mathbf{p}_e\}_{e\in\mathcal{E}}\).
Let
\(\mathcal{I}=\{(e,f,k)\mid e\in\mathcal{E},f\in\mathcal{F}_e,k\in\mathcal{K}_f\}\).
Given \(W_T,W_E\geq 0\) and \(W_T+W_E=1\), the overall problem is formulated as
\begin{subequations}
\label{eq:overall_problem}
\begin{align}
\min_{\boldsymbol{x},\boldsymbol{a}}\quad
& W_T\sum_{e\in\mathcal{E}}\frac{T_e}{T^{\mathrm{ref}}}
+ W_E\sum_{e\in\mathcal{E}}\frac{E_e}{E^{\mathrm{ref}}}
\label{eq:overall_objective}
\\
\mathrm{s.t.}\quad
& x_{f,k}^{O}+x_{f,k}^{T}=1,
\quad \forall (e,f,k)\in\mathcal{I},
\label{eq:unique_assignment}
\\
& r_{f,k}=r_e^{0}+\delta_{i_f,e}^{\mathrm{str}},
\quad \forall (e,f,k)\in\mathcal{I},
\label{eq:release_constraint}
\\
& \sum_{c\in\mathcal{C}}\sum_{j\neq i}
p_{ij,e,h}^{c}\leq P_i^{\max},
\quad \forall e,h,i,
\label{eq:power_constraint}
\\
& \sum_{j\neq i}
\left(
b_{ij,e,h}^{c}+b_{ji,e,h}^{c}
\right)\leq 1,
\quad \forall e,h,i,c.
\label{eq:half_duplex_constraint}
\end{align}
\end{subequations}
Constraint~\eqref{eq:unique_assignment} assigns each chunk exclusively to either the optical or THz fabric. Constraint~\eqref{eq:release_constraint} incorporates the source-rack straggler delay into the chunk release time. Constraint~\eqref{eq:power_constraint} limits the aggregate THz transmit power of each rack during every transmission stage. In \eqref{eq:half_duplex_constraint}, \(b_{ij,e,h}^{c}=1\) indicates that rack \(i\) transmits to rack \(j\) over subband \(c\) during transmission rounds \(h\). This constraint enforces per-subband half-duplex operation by preventing a rack from transmitting and receiving simultaneously.
Problem~\eqref{eq:overall_objective} is a mixed-integer nonlinear program. 
% Its integer variables arise from fabric assignment, collective synthesis, scheduling, and subband selection. 
Directly solving the full problem for every communication event is impractical because the feasible action space grows jointly with the number of flows, chunks, transmission rounds, and subbands. We therefore separate the problem into collective-specific synthesis, wireless resource allocation, and optical-THz coordination developed in Sec. \ref{sec:thz_sync}.

\begin{figure}[!t]
    \centering
    \includegraphics[width=0.8\columnwidth]{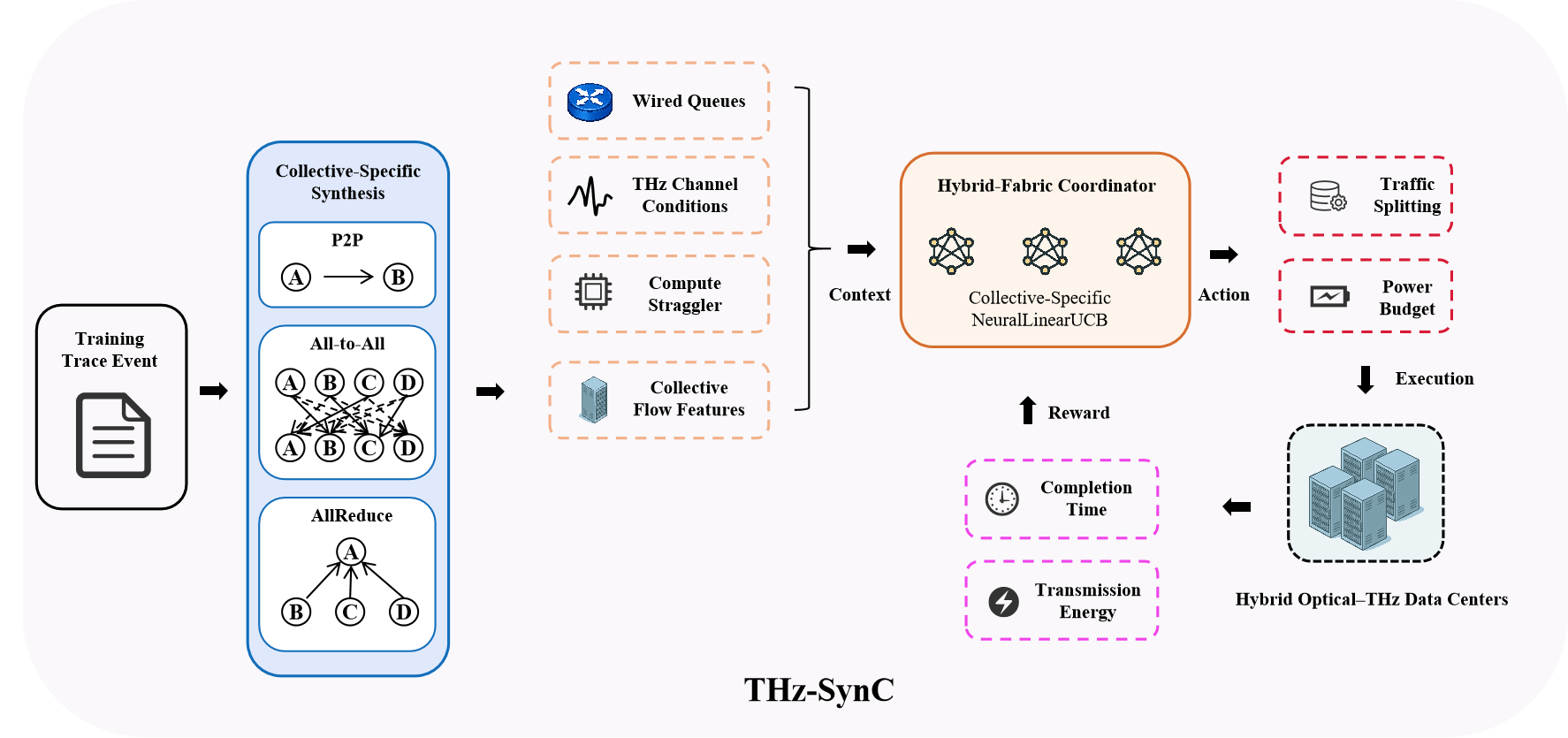}
    \caption{THz-SynC Workflow.}
    \label{fig:thzsync_overview}
\end{figure}

\section{THz-Sync Design}
\label{sec:thz_sync}
\subsection{Overall}
THz-SynC decomposes the joint problem into three layers with distinct responsibilities. Collective synthesis translates a communication event into logical flows and precedence constraints that preserve its semantics. A collective-specific contextual bandit assigns each flow's chunks to the optical or THz fabric and selects a wireless power budget for every transmitting rack. 
The wireless executor then schedules ready THz chunks in synchronous rounds and determines their subbands, transmit powers, and round duration. This separation keeps collective correctness independent of learned decisions while allowing the hybrid fabric to adapt to wired congestion, THz channel fluctuations, and compute stragglers.

The execution workflow of THz-SynC is illustrated in  Fig.~\ref{fig:thzsync_overview}. Once the computational dependencies of an event \(e\) are satisfied, THz-SynC synthesizes its P2P, A2A, or AR flows and their precedence. It then collects the current flow attributes, wired queue states, THz gains, and straggler indicators. The corresponding collective-specific bandit selects the number of THz-owned chunks for every flow and a power budget for each potential transmitting rack. Chunk ownership remains fixed throughout the event. Wired-owned chunks enter the optical network when ready, while THz-owned chunks enter a wireless ready frontier.
The wireless scheduler repeatedly chooses a feasible concurrent set from the frontier. A power allocator computes the minimum powers that serve the selected chunks within a common round duration under the rack budgets. Completion states and collective frontiers are updated synchronously at the end of each round. When all completion conditions of \(e\) hold, its measured completion time and transmission energy produce a delayed reward for the bandit that selected its action. The bandit then updates its action-value estimates using this reward to improve subsequent decisions under similar contexts. Unlike A2A or AR, P2P transmission requires no collective-specific synthesis, and each P2P chunk assigned to the THz fabric is transmitted over the currently best subband using the allocated power budget.

\subsection{All-to-All Synthesis}
\label{sec:a2a_synthesis}

An A2A event specifies a directed demand \(D_{ij}^{e}\) from
each source rack \(i\) to each destination rack \(j\). Its
logical flow set is
\begin{equation}
\mathcal F_e^{\mathrm A}
=
\left\{
f_{ij}
\mid
i,j\in\mathcal V_e,\;
i\neq j,\;
D_{ij}^{e}>0
\right\}.
\label{eq:a2a_flow_set}
\end{equation}
THz-SynC does not impose a predetermined permutation on these transfers. Instead, it executes the event over multiple transmission rounds. At the beginning of round \(h\), the set \(\mathcal A_{e,h}\) contains the THz-owned chunks whose source data have been released and that have not completed.
The scheduler approximates the largest conflict-free subset permitted by the current half-duplex subband and power constraints. 
The selected chunks are transmitted concurrently and complete at the common round boundary. Unselected chunks remain in the frontier, while newly released chunks become eligible in subsequent rounds. Hence, a source straggler delays only its outgoing chunks. This design guarantees complete delivery of the required data while allowing ready sources to proceed independently, thereby mitigating the impact of stragglers on the overall completion time.

For each \(a\in\mathcal A_{e,h}\) and \(c\in\mathcal C\),
let \(b_{a,c}=1\) indicate that chunk \(a\) is transmitted on
subband \(c\). The round schedule is obtained from
\begin{subequations}
\label{eq:a2a_matching}
\begin{align}
\max_{\mathbf b}\quad
& \sum_{a\in\mathcal A_{e,h}}
  \sum_{c\in\mathcal C} b_{a,c}
\label{eq:a2a_matching_obj}
\\
\text{s.t.}\quad
& \sum_{c\in\mathcal C}b_{a,c}\leq1,
&& \forall a\in\mathcal A_{e,h},
\label{eq:a2a_chunk_once}
\\
& \sum_{a:v\in\{i_a,j_a\}}b_{a,c}\leq1,
&& \forall v,\ c\in\mathcal C,
\label{eq:a2a_subband_conflict}
\\
& \sum_{a:v\in\{i_a,j_a\}}
  \sum_{c\in\mathcal C}b_{a,c}
  \leq N_{\mathrm{RF}},
&& \forall v,
\label{eq:a2a_rf_limit}
\\
& b_{a,c}\in\{0,1\},
&& \forall a,\ c.
\label{eq:a2a_matching_binary}
\end{align}
\end{subequations}
% The secondary score ranks equal-cardinality solutions using the observed channel quality. 
Problem~\eqref{eq:a2a_matching} is a generalized
\(b\)-matching problem, where constraint~\eqref{eq:a2a_chunk_once} ensures that each chunk is assigned to at most one subband.
We approximately solve the problem using a channel-aware greedy matching followed by local augmentation.
The matching cardinality is the primary criterion, while the observed channel quality guides the selection of chunk--subband edges. Each candidate edge is inserted only if the subband occupancy remains feasible. The matching is then improved by replacing one selected edge with two feasible unselected edges whenever this exchange increases the number of concurrent transmissions. The resulting chunk-subband assignment is denoted by
\[
\mathcal M_h
=
\left\{
(a,c)\mid b_{a,c}=1
\right\}.
\]

For the fixed assignment \(\mathcal M_h\), THz-SynC
minimizes the common round duration by solving
\begin{equation}
\begin{aligned}
\min_{\Delta_h,\mathbf p}\quad
& \Delta_h
\\
\text{s.t.}\quad
& \sum_{\substack{(a,c)\in\mathcal M_h, i_a=v}}
  p_{a,c}
  \leq P_v^{\mathrm{bud}},
&& \forall v,
\\
& B_{ch}
  \leq
  \Delta_h r_{i_a,j_a}^{c}(p_{a,c}),
&& \forall (a,c)\in\mathcal M_h,
\\
& p_{a,c}\geq0,\quad \Delta_h\geq0.
\end{aligned}
\label{eq:a2a_power_allocation}
\end{equation}
For a trial duration \(\Delta\), the minimum required power
of a selected transmission is
\begin{equation}
p_{a,c}^{\mathrm{req}}(\Delta)
=
\inf
\left\{
p\geq0
\mid
B_{ch}\leq
\Delta r_{i_a,j_a}^{c}(p)
\right\}.
\label{eq:a2a_required_power}
\end{equation}
A trial duration is feasible when
\begin{equation}
\sum_{\substack{(a,c)\in\mathcal M_h, i_a=v}}
p_{a,c}^{\mathrm{req}}(\Delta)
\leq P_v^{\mathrm{bud}},
\qquad \forall v.
\label{eq:a2a_duration_feasibility}
\end{equation}
Since \(p_{a,c}^{\mathrm{req}}(\Delta)\) decreases
monotonically with \(\Delta\), an upper bound is first expanded until feasibility is reached, after which bisection returns \(\Delta_h\) and the corresponding minimum powers.
All chunks in \(\mathcal M_h\) complete at the common round boundary. The ready frontier is then updated, and the procedure repeats until every directed A2A chunk has completed. 
% The execution process of A2A is summarized in Algorithm 1.

% \begin{algorithm}[t]
% \caption{A2A round Synthesis and Execution}
% \label{alg:a2a_synthesis}
% \begin{algorithmic}[1]
% \WHILE{event \(e\) has incomplete chunks}
%     \STATE Update the ready set \(\mathcal A_{e,h}\)
%     \IF{\(\mathcal A_{e,h}=\emptyset\)}
%         \STATE Advance to the next release instant
%     \ELSE
%         \STATE Build all feasible chunk--subband edges
%         \STATE Construct a greedy generalized \(b\)-matching
%         \STATE Improve the matching by local augmentation
%         \STATE Fix \(\mathcal M_h\) and bisect
%         Eq.~\eqref{eq:a2a_power_allocation}
%         \STATE Execute the round and update chunk states
%     \ENDIF
% \ENDWHILE
% \end{algorithmic}
% \end{algorithm}

\subsection{AllReduce Synthesis}

% A conventional ring AR executes reduce-scatter and all-gather over \(2(N-1)\) ordered rounds, where \(N\) is the number of participating racks. Its dependency depth therefore increases linearly with \(N\). Furthermore, a delayed rack can stall the current round and propagate its delay through all subsequent rounds. However, the THz network allows a rack to communicate with multiple neighbors concurrently over different subbands. THz-SynC exploits this capability by partitioning the tensor into multiple shards and assigning each shard to a distinct rooted tree, which enables reduce and broadcast transmissions from different trees to proceed concurrently over distinct subbands. This parallelism reduces the number of sequential communication rounds required. The tree construction further places racks with compute delays closer to the roots, thereby shortening their remaining dependency paths and mitigating the impact of stragglers. For clarity, we consider \(N_c\) trees in the following discussion.
A conventional ring AR performs reduce-scatter and all-gather over $2(N-1)$ ordered rounds, giving a dependency depth that grows linearly with the number of participating racks $N$. A delayed rack can therefore stall the current round and propagate its delay to subsequent rounds. In contrast, THz links allow each rack to communicate with multiple neighbors concurrently on different subbands. THz-SynC partitions the tensor into multiple shards and assigns each shard to a distinct rooted tree, enabling reduce and broadcast transmissions across trees to proceed in parallel over separate subbands. This reduces sequential communication rounds. The tree construction also places compute-delayed racks closer to the roots, shortening their remaining dependency paths and mitigating straggler impact. We consider $N_c$ trees below.

Let \(D_e^{\mathrm R}\) be the tensor size of AR event \(e\). It is divided into \(N_c\) shards indexed by \(\mathcal S=\{1,2,\dots,N_c\}\).
Shard \(\sigma\) is carried by a directed tree
\(\mathcal T_\sigma=(\mathcal V_e,\mathcal E_\sigma^{\mathrm T})\) with root \(o_\sigma\). 
Reduce transmissions follow child-to-parent edges, while broadcast transmissions reverse these edges.
At event activation, THz-SynC selects \(N_c\) distinct roots from the participating racks. Racks with larger delays are preferred as roots when their THz links to the remaining participants are sufficiently reliable, which places potential stragglers closer to the aggregation endpoints.
The remaining roots are selected from racks that maintain generally favorable channel conditions to the other participants.
Each tree is expanded greedily under the common fanout limit \(N_c\). Racks with larger straggler delays are inserted first, followed by racks with better channel conditions. For each new rack, feasible parents are the racks already inserted in
the tree that have residual fanout to the new rack.

For shard \(\sigma\) and node \(v\), let \(r_{\sigma,v}^{\mathrm{loc}}\) denote the availability time of the local contribution. A non-root node can transmit a reduced chunk only after its
local contribution and the corresponding contributions from
all children are available. Its reduce-ready time is
\begin{equation}
R_{\sigma,v}^{\mathrm{red}}
=
\max\!\left\{
R_{\sigma,v}^{\mathrm{loc}},
\max_{u\in\operatorname{ch}_{\sigma}(v)}
C_{\sigma,u}^{\mathrm{red}}
\right\}.
\label{eq:ar_reduce_release}
\end{equation}
For a leaf node, the second term is omitted. This dependency allows ready subtrees to proceed independently, while a delayed rack affects only the path from that rack to the root.
The reduction of shard \(\sigma\) completes at its root at
\begin{equation}
J_{\sigma}
=
\max\!\left\{
R_{\sigma,o_{\sigma}}^{\mathrm{loc}},
\max_{u\in\operatorname{ch}_{\sigma}(o_{\sigma})}
C_{\sigma,u}^{\mathrm{red}}
\right\}.
\label{eq:ar_root_join}
\end{equation}
The root starts broadcasting after \(J_{\sigma}\). Each non-root node can forward a broadcast chunk after receiving that chunk from its parent, which can be stated as
\begin{equation}
R_{\sigma,v}^{\mathrm{b}}
=
\begin{cases}
J_{\sigma},
& v=o_{\sigma},\\
C_{\sigma,v}^{\mathrm{b}},
& v\neq o_{\sigma}.
\end{cases}
\label{eq:ar_broadcast_release}
\end{equation}
Event \(e\) completes only after every broadcast chunk of all shards has arrived.

Unlike A2A, the transmission order of AR chunks is constrained by the tree dependencies. 
In each transmission round, all currently ready child-to-parent transmissions in the reduce
phase, or parent-to-child transmissions in the broadcast phase, are executed concurrently subject to the wireless resource constraints.
Let \(\mathcal M_h\) still denote the transmissions assigned to wireless subround \(h\). Since their logical order has already been determined by the precedence executor, the remaining problem assigns one subband to every transmission and minimizes the common subround duration as follows
\begin{subequations}
\label{eq:ar_round_problem}
\begin{align}
\min_{\Delta_h,\mathbf b,\mathbf p}\quad
& \Delta_h
\label{eq:ar_round_obj}
\\
\text{s.t.}\quad
& \sum_{c\in\mathcal C}b_{a,c}=1,
&& \forall a\in\mathcal M_h,
\label{eq:ar_one_subband}
\\
& \sum_{\substack{a\in\mathcal M_h, 
v\in\{i_a,j_a\}}}b_{a,c}\leq1,
&& \forall v,\ c\in\mathcal C,
\label{eq:ar_subband_conflict}
\\
& \sum_{\substack{a\in\mathcal M_h, v\in\{i_a,j_a\}}}
  \sum_{c\in\mathcal C}b_{a,c}
  \leq N_c,
&& \forall v,
\label{eq:ar_rf_limit}
\\
& \sum_{\substack{a\in\mathcal M_h, i_a=v}}
  \sum_{c\in\mathcal C}p_{a,c}
  \leq P_v^{\mathrm{bud}},
&& \forall v,
\label{eq:ar_power_budget}
\\
& B_a b_{a,c}
  \leq
  \Delta_h r_{i_a,j_a}^{c}(p_{a,c}),
&& \forall a,\ c,
\label{eq:ar_round_service}
\\
& b_{a,c}\in\{0,1\},
\quad \Delta_h\geq0.
\label{eq:ar_round_domain}
\end{align}
\end{subequations}
Here, \(B_a\) denotes the data volume of all
THz-owned chunks carried by tree-edge transmission \(a\).

THz-SynC uses a channel-aware greedy procedure for subband assignment. Ready transmissions with fewer suitable subbands or weaker channel conditions are processed first. Each transmission is assigned the available subband with the highest channel gain that satisfies the endpoint and subband constraints.
After the subband assignment is done, the continuous power allocator determines the minimum feasible subround duration and the corresponding transmit powers.

For the resulting subband mapping \(c(a)\), the minimum power required by transmission \(a\) under a trial duration
\(\Delta\) is
\begin{equation}
p_a^{\mathrm{req}}(\Delta)
=
\inf\left\{
p\geq0
\;\middle|\;
B_a\leq
\Delta r_{i_a,j_a}^{c(a)}(p)
\right\}.
\label{eq:ar_required_power}
\end{equation}
The trial duration is feasible if
\begin{equation}
\sum_{\substack{a\in\mathcal M_h, i_a=v}}
p_a^{\mathrm{req}}(\Delta)
\leq
P_v^{\mathrm{bud}},
\qquad \forall v.
\label{eq:ar_power_feasibility}
\end{equation}
Because \(p_a^{\mathrm{req}}(\Delta)\) decreases monotonically with \(\Delta\), the executor expands an initial upper bound until feasibility is reached and then applies bisection. The returned powers are
\begin{equation}
p_a^\star
=
p_a^{\mathrm{req}}(\Delta_h^\star).
\label{eq:ar_optimal_power}
\end{equation}
This procedure obtains the minimum common duration for the given subband assignment within the bisection
tolerance. 
% The AR execution process is summarized in Algorithm 2.

% \begin{algorithm}[t]
% \caption{\(N_c\)-Tree AllReduce Synthesis and Execution}
% \label{alg:allreduce_synthesis}
% \begin{algorithmic}[1]
% \REQUIRE Event \(e\), racks \(\mathcal V_e\), tensor
% \(D_e^{\mathrm R}\), channel state, and
% \(\{P_v^{\mathrm{bud}}\}\)
% \STATE Divide \(D_e^{\mathrm R}\) into \(N_c\) shards
% \STATE Select \(N_c\) distinct roots according to rack delays
% and channel conditions
% \FOR{\(\sigma\in\{1,\ldots,N_c\}\)}
%     \STATE Initialize \(\mathcal T_\sigma\) with root
%     \(o_\sigma\)
%     \STATE Order the remaining racks by delay and channel
%     condition
%     \FOR{each rack \(v\) in the order}
%         \STATE Select a parent with residual fanout and a
%         reliable link to \(v\)
%         \STATE Add \(v\) and its parent edge to
%         \(\mathcal T_\sigma\)
%     \ENDFOR
% \ENDFOR
% \STATE Establish reduce and broadcast dependencies
% \WHILE{event \(e\) is incomplete}
%     \STATE Obtain the currently ready tree-edge transmissions
%     \STATE Order ready transmissions by subband availability and channel quality
%     \STATE Greedily assign each transmission its highest-gain feasible subband
%     \STATE Expand the duration bound until
%     \eqref{eq:ar_power_feasibility} is satisfied
%     \STATE Bisect the bound to obtain
%     \(\Delta_h^\star\) and
%     \(p_a^\star=p_a^{\mathrm{req}}(\Delta_h^\star)\)
%     \STATE Execute and update the transmission dependencies
% \ENDWHILE
% \end{algorithmic}
% \end{algorithm}

\subsection{Collective-Specific Contextual Bandit Design}

The performance of the optical and THz fabrics varies with the current wired queues, THz channel conditions, flow sizes, and compute delays. A fixed traffic split rule cannot adapt to these variations. THz-SynC therefore uses a contextual bandit to learn the performance of fabric coordination actions from previously completed collective events.
Moreover, a contextual bandit is well suited to this problem because the action selected for one collective does not determine the logical state of the next collective through a reusable Markov transition.
Although network queues and channels evolve, their current values are directly included in the context observed at each decision. This design avoids learning unnecessary state transitions and assigns one delayed outcome to each complete coordination action.
THz-SynC maintains an independent NeuralLinearUCB model for P2P, A2A, and AR. This separation allows each model to learn the completion behavior of its own flow structure and collective semantics without interference from the reward distributions of the other collective types.

Let \(\boldsymbol{y}_e\) denote the context observed when collective event \(e\) becomes active. For each flow \(f\in\mathcal F_e\), the current implementation constructs the feature vector as
\begin{equation}
\boldsymbol{y}_{e,f}
=
\left[
i_f,\,
j_f,\,
M_f,\,
s_{i_f,e},\,
Q_{i_fj_f},\,
U_{i_fj_f},\,
\left\{ g_{i_fj_f}^{c}\right\}_{c\in\mathcal C}
\right].
\label{eq:bandit_flow_context}
\end{equation}
Here, \(i_f\) and \(j_f\) are the source and destination rack indices, and \(\tilde M_f\)
represents the number of chunks carried by the
flow. The binary variable \(s_{i_f,e}\) indicates whether the source rack experiences a compute delay. The terms \(Q_{i_fj_f}\) and \( U_{i_fj_f}\) denote the queued data and the maximum shared-buffer
occupancy along the wired path, respectively.
The remaining entries are the normalized THz gains on the \(N_c\) subbands.

The bandit selects one action when event \(e\) is activated
\begin{equation}
\boldsymbol{a}_e
=
\left(
\left\{n_{e,f}^{H}\right\}_{f\in\mathcal F_e},
\left\{P_{e,i}^{\mathrm{bud}}\right\}_{i\in\mathcal V_e}
\right).
\label{eq:bandit_action}
\end{equation}
For a flow containing \(K_f\) chunks,
\(n_{e,f}^{H}\in\{0,\ldots,K_f\}\) specifies the number of chunks assigned to the THz fabric. The second action component specifies the power budget of each sending rack \(P_{e,i}^{\mathrm{bud}} = \beta_{e,i}P_i^{\max}, \qquad \beta_{e,i}\in\mathcal B\). \(\mathcal B\) is the configured set of normalized power levels. A rack with no THz-owned outgoing chunk is assigned zero power. 
The complete Cartesian product of the flow counts and rack power levels grows rapidly with the number of flows. Therefore, THz-SynC evaluates a fixed-size structured candidate set \(\mathcal A_e\), rather than enumerating the complete action space. Each action is encoded as
\begin{equation}
\boldsymbol{v}_e(\boldsymbol a)
=
\left[
\left\{
\frac{n_{e,f}^{H}}{M_f}
\right\}_{f\in\mathcal F_e},
\left\{
\frac{P_{e,i}^{\mathrm{bud}}}{P_i^{\max}}
\right\}_{i\in\mathcal V_e}
\right],
\label{eq:bandit_action_vector}
\end{equation}
with the same padding and masks as the context.

The joint action is generated through a coordinate-wise beam search. The search progressively expands the THz chunk count of each flow and the power level of each active sender, while retaining only the highest-scoring partial actions at every step until a complete action is obtained. 

For each collective type, NeuralLinearUCB maps the concatenated context and action vector to a nonlinear representation, which can be described as
\begin{equation}
\boldsymbol{\phi}_{e,a}
=
\phi_{\theta}
\left(
\left[
\boldsymbol{y}_e,
\boldsymbol{v}_e(\boldsymbol a)
\right]
\right).
\label{eq:bandit_representation}
\end{equation}
The linear head estimates the expected reward and its
uncertainty as
\begin{align}
\widehat{\mu}_{e}(\boldsymbol a)
&=
\widehat{\boldsymbol w}^{\mathsf T}
\boldsymbol{\phi}_{e,a},
\label{eq:bandit_mean}
\\
\sigma_e(\boldsymbol a)
&=
\sqrt{
\boldsymbol{\phi}_{e,a}^{\mathsf T}
\boldsymbol A^{-1}
\boldsymbol{\phi}_{e,a}
}.
\label{eq:bandit_uncertainty}
\end{align}
The selected action is
\begin{equation}
\boldsymbol a_e^\star
=
\arg\max_{\boldsymbol a\in\mathcal A_e}
\left[
\widehat{\mu}_{e}(\boldsymbol a)
+
\alpha\sigma_e(\boldsymbol a)
\right],
\label{eq:bandit_ucb_selection}
\end{equation}
where \(\alpha\) controls the exploration of uncertain actions. The selected ownership and rack power budgets are then committed for the complete event.

After event \(e\) completes, THz-SynC computes the delayed reward
\(r_e
=
-
\left(
W_T
\frac{T_e}{T_{\tau(e)}^{\mathrm{ref}}}
+
W_E
\frac{E_e^{W}+E_e^{H}}
     {E_{\tau(e)}^{\mathrm{ref}}}
\right).\)
Each event produces one reward for its complete action. For the completed sample, the linear posterior is updated as
\begin{align}
\boldsymbol A
&\leftarrow
\boldsymbol A+
\boldsymbol{\phi}_{e,a^\star}
\boldsymbol{\phi}_{e,a^\star}^{\mathsf T},
\label{eq:bandit_A_update}
\\
\boldsymbol b
&\leftarrow
\boldsymbol b+
r_e\boldsymbol{\phi}_{e,a^\star},
\qquad
\widehat{\boldsymbol w}
=
\boldsymbol A^{-1}\boldsymbol b.
\label{eq:bandit_b_update}
\end{align}
The full context-action sample and reward are also appended to the replay buffer of the corresponding collective. At fixed update intervals, the neural encoder is trained from replay samples using the squared reward-prediction error. Since the feature representation changes after encoder training, the
linear posterior is subsequently rebuilt from the replay buffer. This procedure enables immediate online updates while allowing the nonlinear representation to adapt to accumulated network observations.

\section{Performance Evaluation}

\subsection{Simulation Setup}
We develop an event-driven simulator that jointly models the training dependency graph, the shared wired optical fabric, and the reconfigurable THz overlay. Unless otherwise specified, the simulated partition contains 32 racks arranged on four concentric rings, with eight racks per ring and one ring assigned to each PP stage. The innermost-ring radius is $\rho_0=4~\mathrm{m}$, and the inter-ring spacing is $\Delta\rho=2~\mathrm{m}$. Both the rack-to-switch access links and the inter-ring switch links operate at $C_{\mathrm{acc}}=C_{\mathrm{inter}}=100~\mathrm{Gbps}$. The wired transmission-energy coefficient is $\epsilon^{O}=40~\mathrm{pJ/bit/hop}$. The THz overlay provides $N_c=4$ non-overlapping subbands, each with bandwidth $B_c=5~\mathrm{GHz}$, spanning the $290$--$310~\mathrm{GHz}$ band. We set
the receive-antenna gain to $G=25~\mathrm{dBi}$ and the maximum transmit power of each rack to $P_i^{\max}=0.1~\mathrm{W}$. Each inter-rack flow is partitioned into scheduling chunks of at most
$B_{\mathrm{ch}}=512~\mathrm{KiB}$.

Our workload generator follows the dependency structure of the non-interleaved 1F1B pipeline schedule in Megatron-LM. We use four PP stages, eight data-parallel ranks per stage, and eight microbatches per iteration. MoE layers are placed in two PP stages, and gradient synchronization is divided into four AR buckets. Rather than imposing a predefined communication mixture, the generator releases forward-activation and backward-gradient P2P events, MoE dispatch and combine A2A events, and AR
events according to their producer dependencies in the training graph. Each P2P flow carries $16~\mathrm{MiB}$, each source contributes $12~\mathrm{MiB}$ to an A2A operation, and each AR bucket contains $128~\mathrm{MiB}$. For each experiment, all compared methods replay the same training trace and THz channel realizations.

% \begin{figure}[!t]
%     \centering
%     \includegraphics[width=0.9\columnwidth]{fig/reward_vs_iteration.png}
%     \caption{Reward v.s. Iteration.}
%     \label{fig:reward}
% \end{figure}

% \begin{figure}[!t]
%     \centering
%     \includegraphics[width=0.9\columnwidth]{fig/energy_vs_iteration.png}
%     \caption{Total Transmission Energy v.s. Iteration.}
%     \label{fig:energy}
% \end{figure}

% \begin{figure}[!t]
%     \centering
%     \includegraphics[width=0.9\columnwidth]{fig/latency_vs_iteration.png}
%     \caption{Total Collective Completion Time v.s. Iteration.}
%     \label{fig:latency}
% \end{figure}

\begin{figure*}[!t]
    \centering

    \begin{minipage}[t]{0.32\textwidth}
        \centering
        \vspace{0pt}
        \includegraphics[width=\linewidth]
        {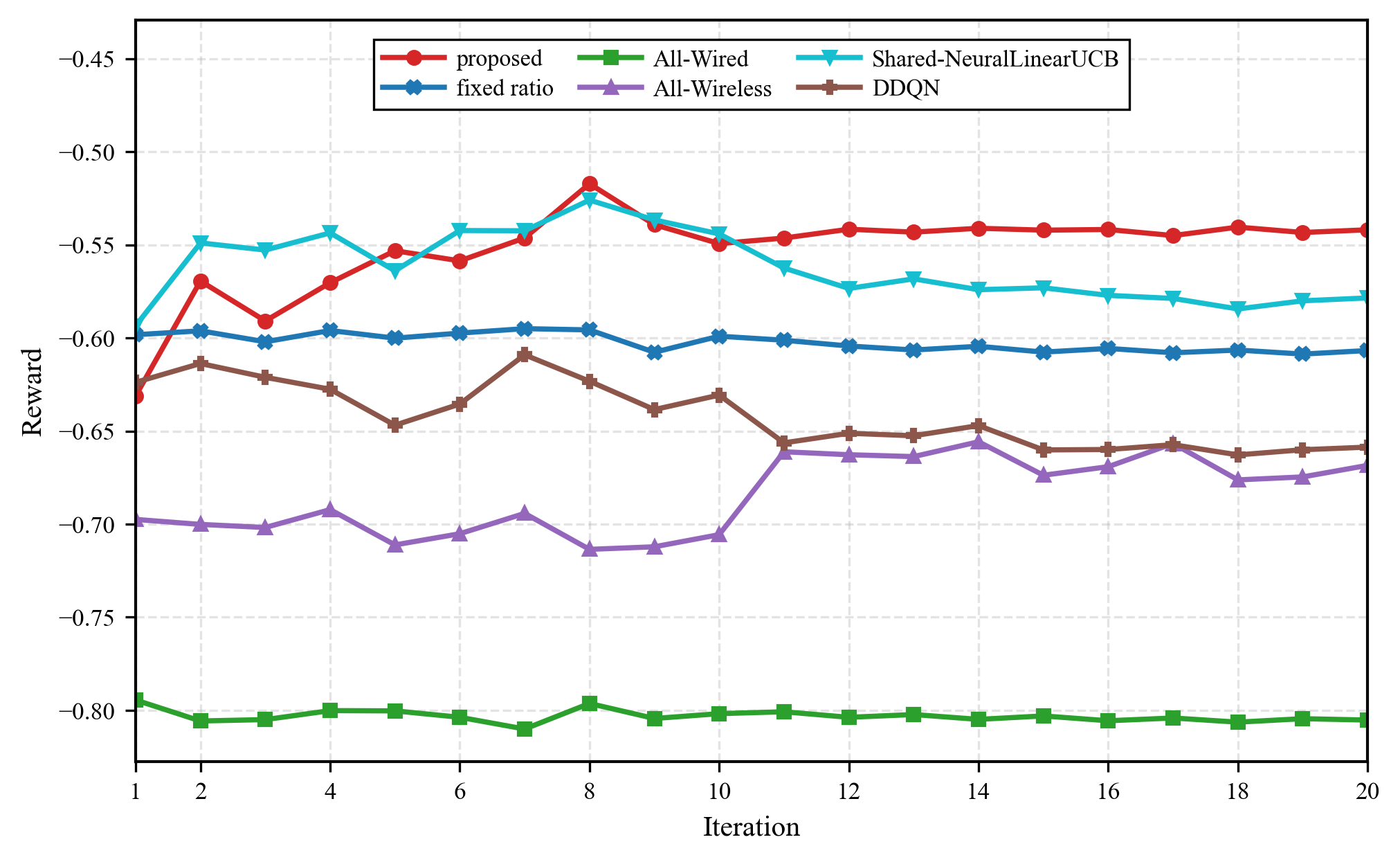}
        \caption{Reward versus iteration.}
        \label{fig:reward}
    \end{minipage}
    \hfill
    \begin{minipage}[t]{0.32\textwidth}
        \centering
        \vspace{0pt}
        \includegraphics[width=\linewidth]
        {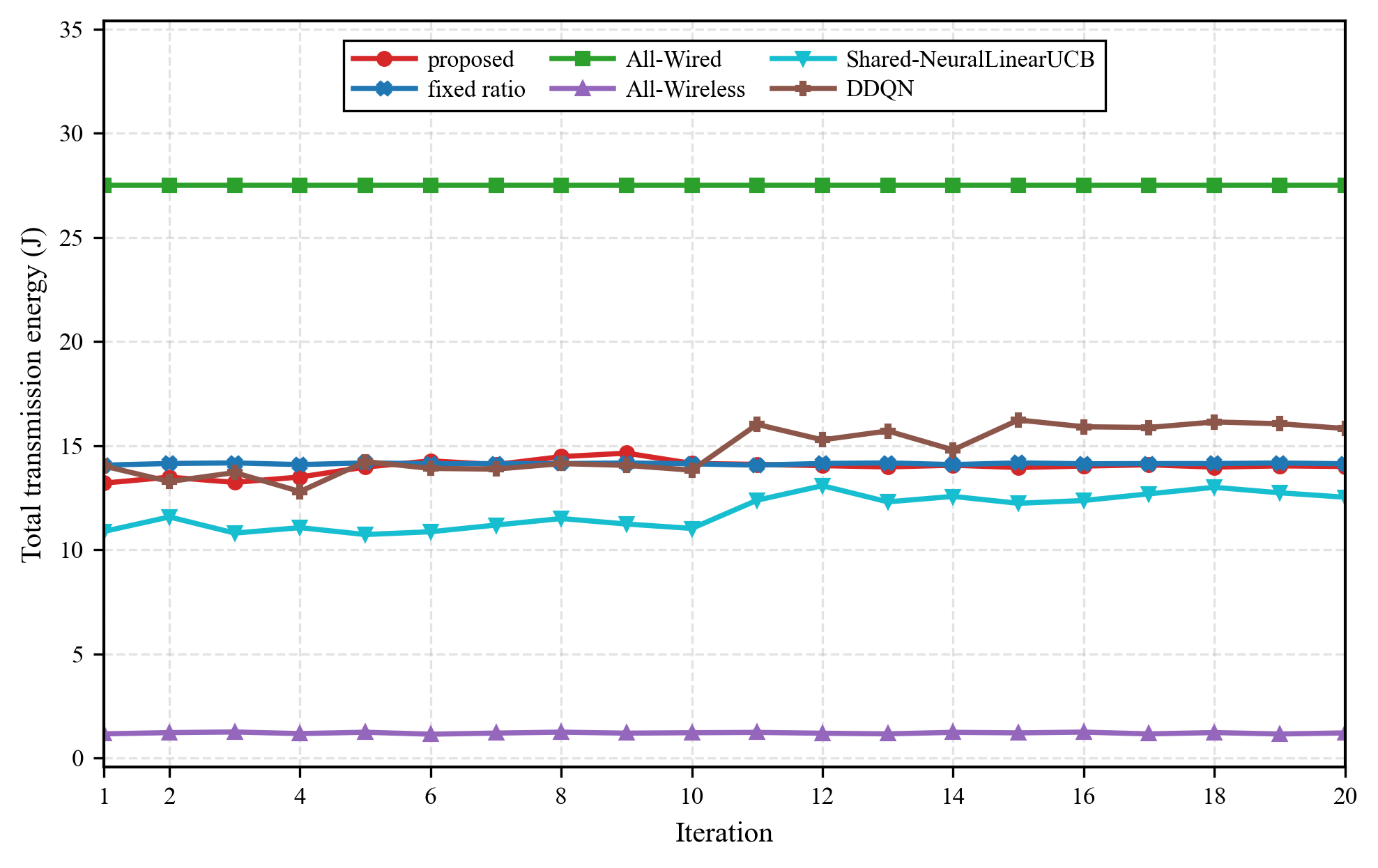}
        \caption{Total transmission energy versus iteration.}
        \label{fig:energy}
    \end{minipage}
    \hfill
    \begin{minipage}[t]{0.32\textwidth}
        \centering
        \vspace{0pt}
        \includegraphics[width=\linewidth]
        {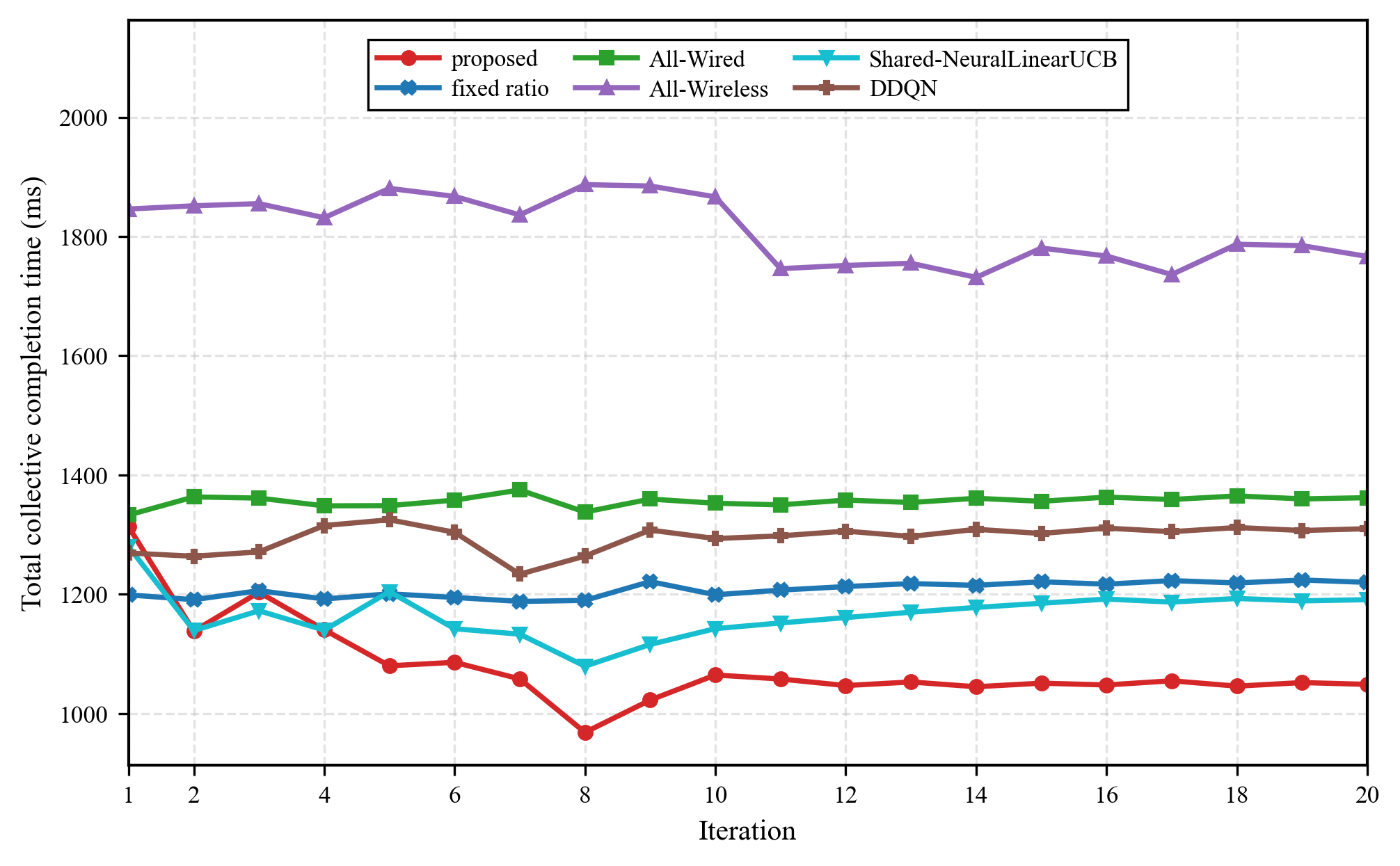}
        \caption{Total collective completion time versus iteration.}
        \label{fig:latency}
    \end{minipage}

\end{figure*}

\begin{figure*}[!t]
    \centering
    \subfloat[A2A with compute stragglers.]{
        \includegraphics[width=0.235\textwidth]
        {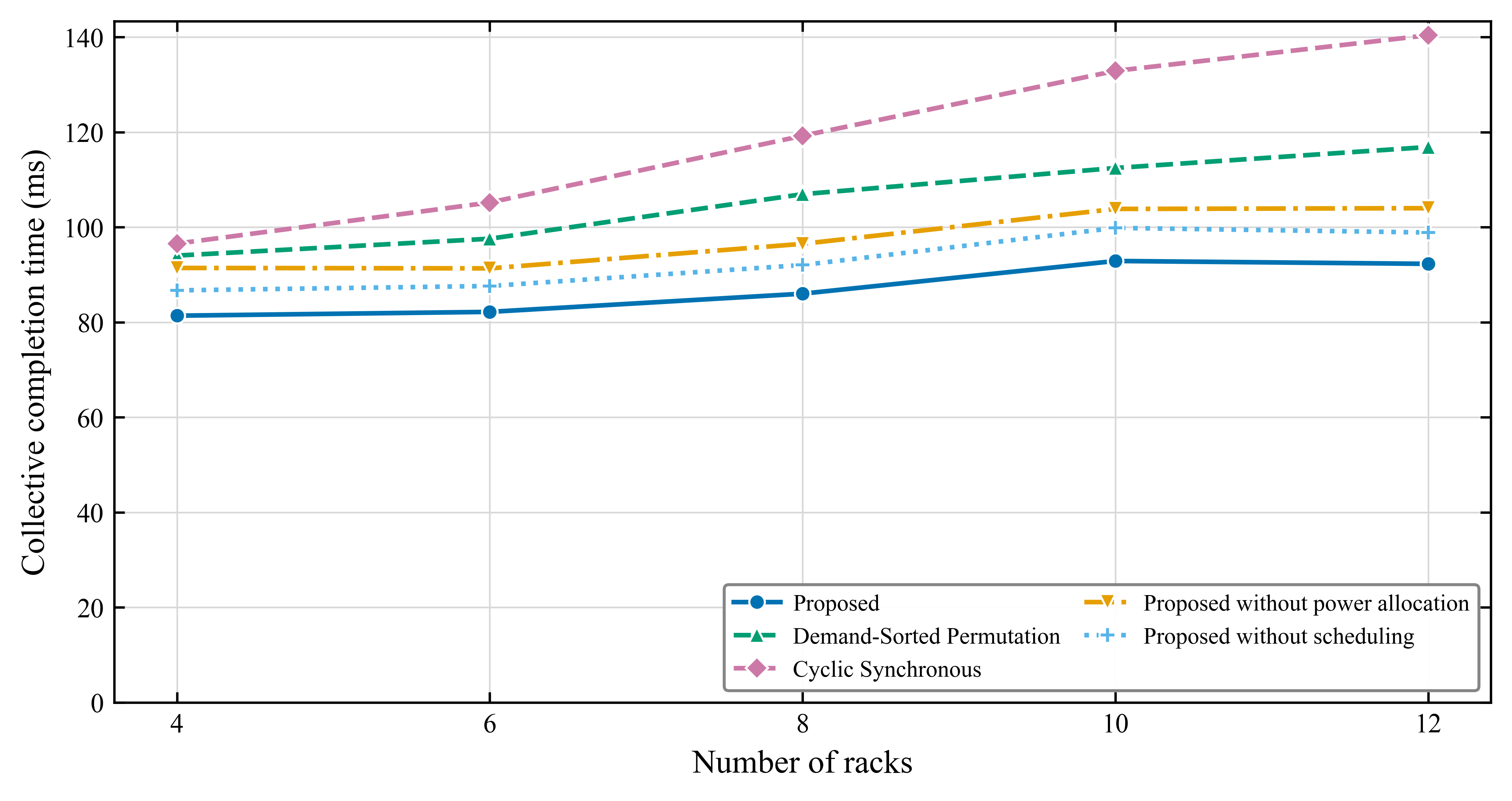}
        \label{fig:a2a_straggler}
    }
    \hfill
    \subfloat[A2A without compute stragglers.]{
        \includegraphics[width=0.235\textwidth]
        {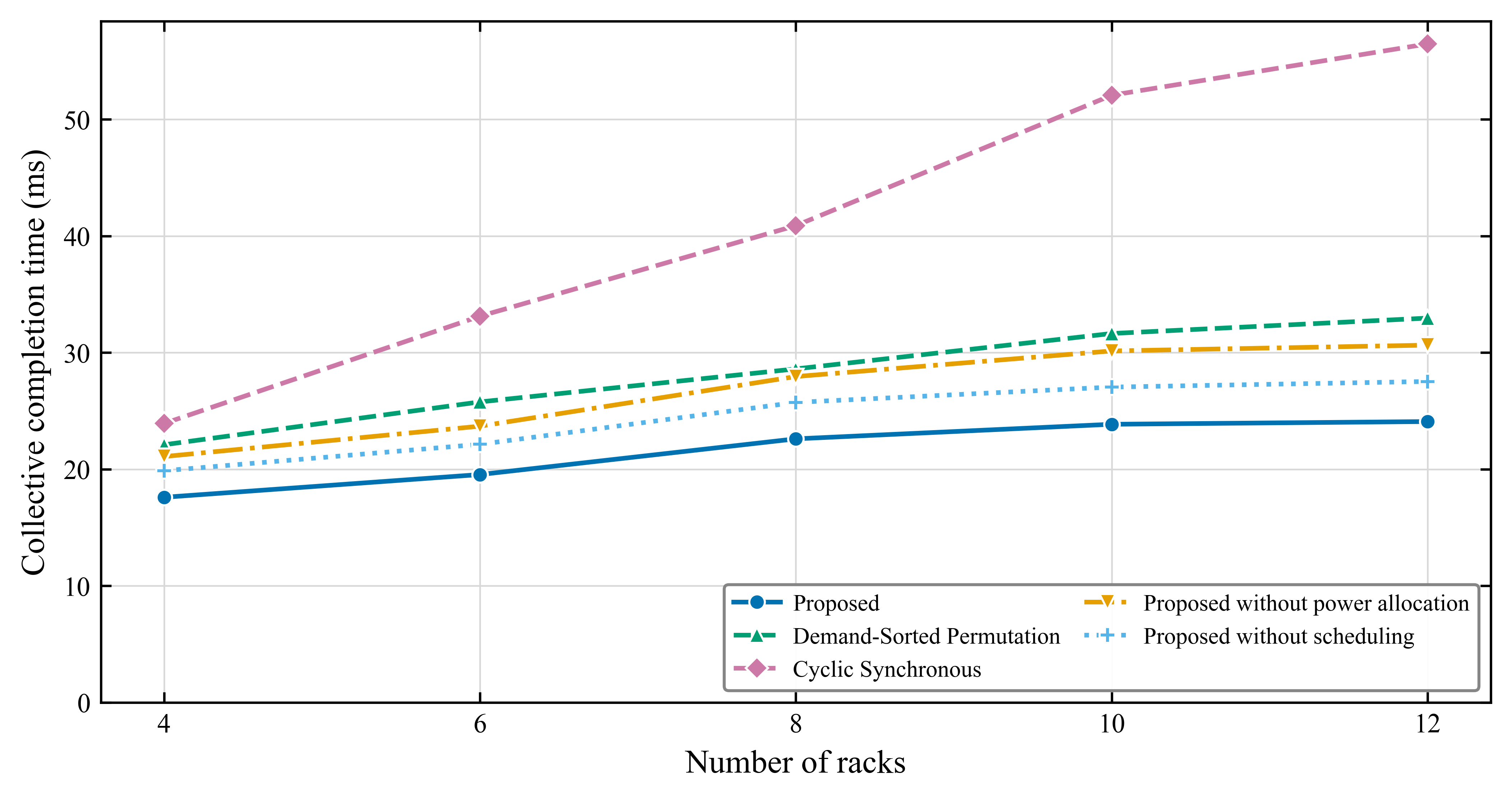}
        \label{fig:a2a_no_straggler}
    }
    \hfill
    \subfloat[AR with compute stragglers.]{
        \includegraphics[width=0.235\textwidth]
        {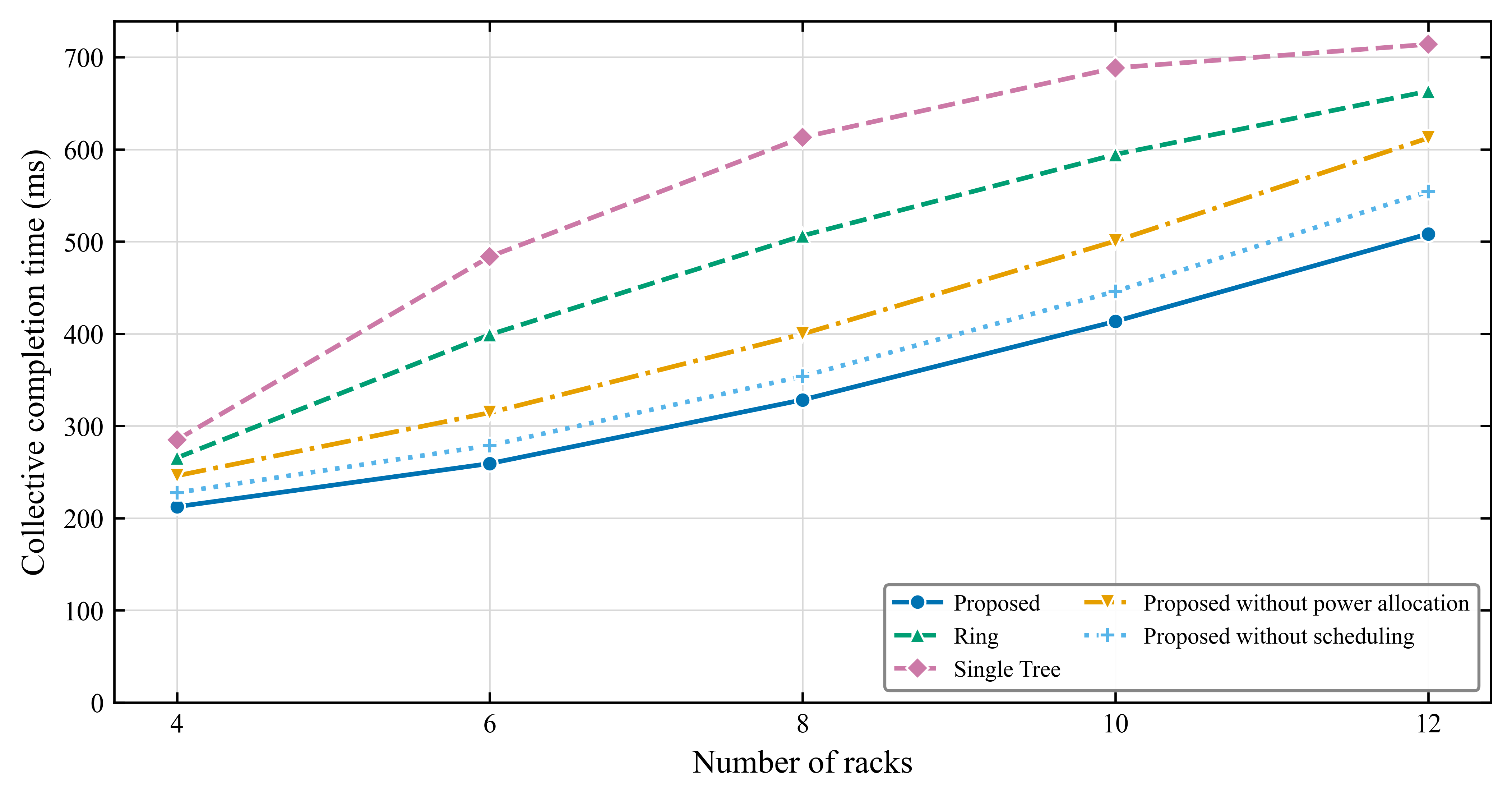}
        \label{fig:allreduce_straggler}
    }
    \hfill
    \subfloat[AR without compute stragglers.]{
        \includegraphics[width=0.235\textwidth]
        {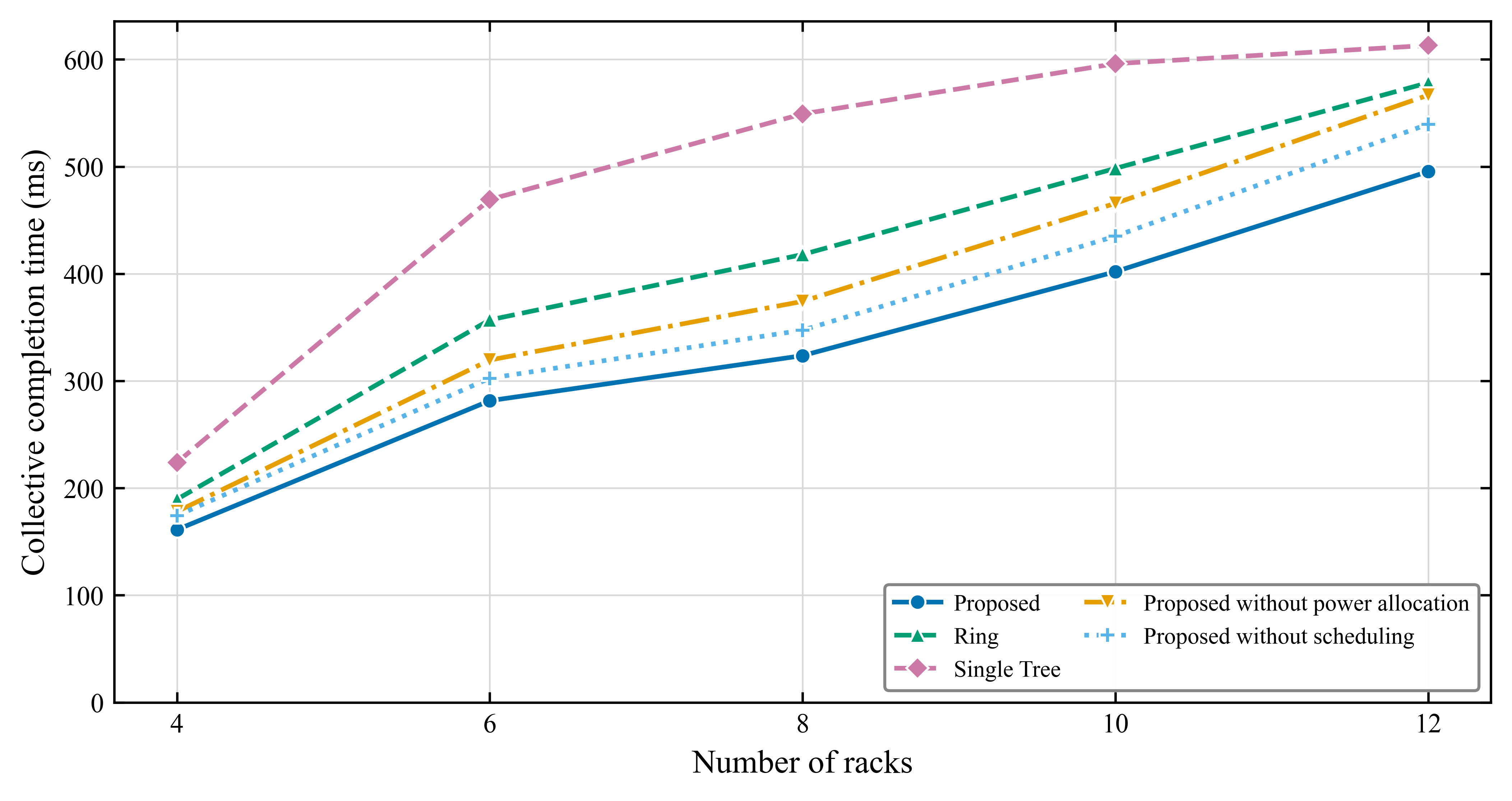}
        \label{fig:allreduce_no_straggler}
    }
    \caption{Completion-time comparison of different schemes for A2A
    and AR collectives with and without compute stragglers.}
    \label{fig:collective_synthesis}
\end{figure*}

\subsection{Performance of THz-SynC}
We first evaluate whether context-dependent hybrid-fabric decisions improve the delay-energy tradeoff over static fabric assignments and other learning methods. We report the reward, total transmission energy, and total collective completion time in each iteration. The reward weights are $W_T=0.7$ and $W_E=0.3$. We compare THz-SynC with five baselines. \emph{All-Wired} assigns every chunk
to the wired fabric, while \emph{All-Wireless} assigns every chunk to the THz overlay. \emph{Fixed Ratio} assigns 50\% of every flow's chunks to each fabric and uses 50\% of the rack maximum power for active wireless senders. \emph{Shared-NeuralLinearUCB} uses one NeuralLinearUCB model across P2P, A2A, and AR. \emph{DDQN} is a well-established deep reinforcement learning algorithm using the same state and action spaces. 

Figs.\ref{fig:reward}-\ref{fig:latency} compare the performance over 20 training iterations. Averaged over iterations 11-20, THz-SynC achieves a collective completion time of 1,050~ms, an energy consumption of 14.51~J, and the highest reward of $-0.54$. Its latency is 11.3\% lower than \emph{Shared-NeuralLinearUCB}, 13.8\% lower than Fixed Ratio, and 19.6\% lower than \emph{DDQN}. The single-fabric baselines represent two performance extremes. \emph{All-Wired} maintains a completion time of 1,359~ms but consumes 27.52~J, whereas \emph{All-Wireless} reduces the energy to 1.13~J at the cost of increasing the completion time to 1,758~ms. \emph{Shared-NeuralLinearUCB} consumes slightly less energy than THz-SynC, at 12.71~J, but its completion time increases to 1,186~ms.

The low energy consumption of All-Wireless confirms that short-range THz transmission can be more energy-efficient than a switched wired fabric in the considered data-center topology, which is consistent with analysis in \cite{han2026wires}. However, relying exclusively on either fabric cannot jointly address network fluctuations. THz-SynC instead learns these dynamics and adapts the fabric split and power budget for each communication event, outperforming the static single-fabric and fixed-ratio policies. It also achieves higher reward than DDQN and the shared-bandit variant. The former is less effective in this one-step decision problem, while the latter uses a common model across heterogeneous collective semantics. Assigning a dedicated bandit to each collective avoids such cross-semantic interference. Since the current objective places greater weight on delay, THz-SynC does not minimize energy alone. It accepts a moderate energy increase over Shared-NeuralLinearUCB to obtain the lowest completion time and, consequently, the highest overall reward.

\subsection{Performance of Collective Synthesis}
We next isolate the effect of collective synthesis from the long-term training trace and the hybrid-fabric controller. In this experiment, we only consider a single THz wireless ring and do not include the wired fabric, background traffic, or bandit decisions.
The physical ring has 16 fixed rack positions and a diameter of 20\,m. For a scale of $N\in\ 4,6,8,10,12\}$, we activate the first $N$ positions instead of redistributing the racks over the entire
ring. Each point in Fig.~\ref{fig:collective_synthesis} reports the mean completion time over 30 paired Monte Carlo seeds.  In the no-straggler case, the data generated by all racks become ready simultaneously at the nominal release time. In the straggler case, $\max(1,\lceil N/8\rceil)$ racks receive an additional arrival delay uniformly sampled between 50 and 100 ms. 
We use a $512~\mathrm{MiB}$ tensor for AR and generate a directed A2A demand matrix with exactly $64~\mathrm{MiB}$ of outgoing traffic per rack.

Figures~\ref{fig:collective_synthesis}(a) and \ref{fig:collective_synthesis}(b) evaluate A2A synthesis. 
We compare emph{Proposed} with four baselines. \emph{Demand-Sorted Permutation} repeatedly constructs one-to-one permutations by prioritizing the greatest remaining source-destination demands and completes each permutation before forming the next one. \emph{Cyclic Synchronous} instead serves rack pairs according to fixed cyclic offsets and advances to the next offset only after all transfers in the current offset complete. The remaining two baselines are ablations of \emph{Proposed}. Without stragglers, the completion time of \emph{Proposed} increases from 17.7~ms at four racks to 24.2~ms at 12 racks. At $N=12$, it reduces completion time by 26.2\% over \emph{Demand-Sorted Permutation} and 57.1\% over \emph{Cyclic Synchronous}. Under stragglers, \emph{Proposed} still completes in 92.3~ms, compared with 116.2 and 140.4~ms for the two baselines. 

Figures~\ref{fig:collective_synthesis}(c) and \ref{fig:collective_synthesis}(d) report the AR results.
\emph{Proposed} executes $N_c$ rooted trees concurrently, while \emph{Ring} retains $2(N-1)$ dependent stages and \emph{Single Tree} uses one unsharded reduction tree. At $N=12$, \emph{Proposed} completes in
495.4~ms without stragglers, compared with 577.2~ms for \emph{Ring} and 613.0~ms for \emph{Single Tree}. With stragglers, the corresponding times are 507.9, 659.0, and 714.2~ms, yielding reductions of 22.9\% and 28.9\%. The additional straggler-induced delay is only 12.5~ms for \emph{Proposed}, but reaches 81.8~ms for \emph{Ring} and 101.2~ms for \emph{Single Tree}. Concurrent trees and readiness-aware dependencies therefore limit the extent to which a delayed rack propagates along the collective critical path.

The ablations further confirm the contributions of both physical scheduling and power allocation. At $N=12$, replacing channel-aware scheduling increases completion time by 7.3--14.0\% across the four evaluated A2A and AR settings, while equal power allocation incurs a 13.3--28.1\% increase.
The performance gaps are relatively limited at small rack counts, where contention and dependency depth remain low. As the collective scale grows, synchronization barriers, unfavorable subband assignments, and bottleneck links accumulate, making the benefits of collective-specific synthesis and resource allocation increasingly evident.

\begin{figure}[!t]
    \centering
    \includegraphics[width=0.8\columnwidth]{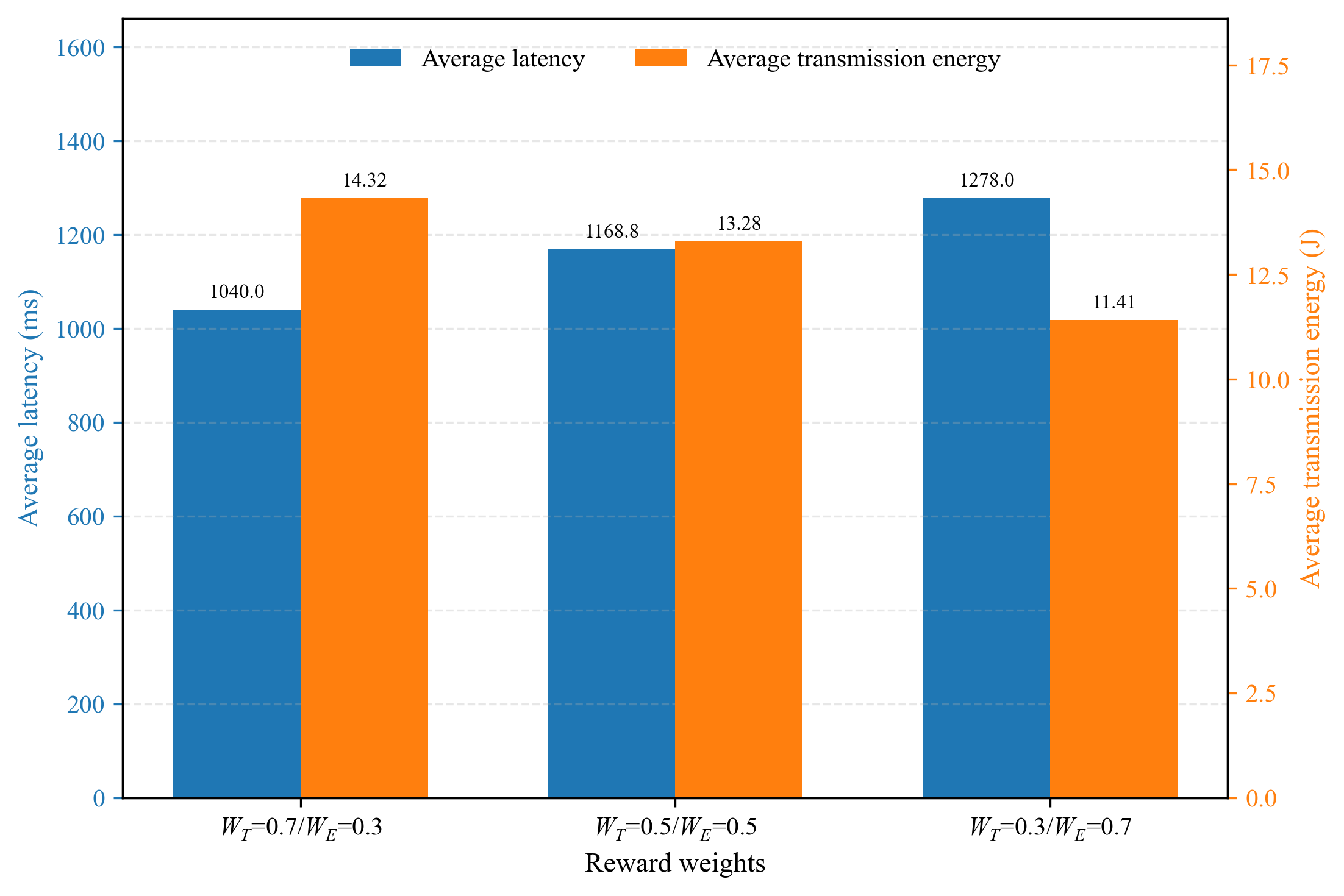}
    \caption{Impact of Reward Weights on the Delay–Energy Tradeoff.}
    \label{fig:reward_weight}
\end{figure}

\subsection{Impact of reward weight}
Finally, we examine how the reward weights control the operating point of THz-SynC. We evaluate three configurations, $(W_T,W_E)\in\{(0.7,0.3),(0.5,0.5),(0.3,0.7)\}$, under the same trace and channel-generation rules. Each bar in Fig.~\ref{fig:reward_weight} reports the average over converged 5 iterations. Increasing the energy weight yields a clear delay–energy tradeoff. Across the weight settings from $(0.7,0.3)$ to $(0.3,0.7)$, latency increases from 1,040.0 to 1,278.0~ms, while energy decreases from 14.32 to 11.41~J, respectively. These results confirm that the reward weights effectively control the operating point, enabling THz-SynC to achieve different tradeoffs according to application requirements.

\section{Conclusion}

We presented THz-SynC, a collective-aware framework for coordinating optical and THz fabrics in distributed AI training data centers. THz-SynC combines collective-specific synthesis for A2A and AR with contextual-bandit-assisted traffic splitting and rack-level power control. Its wireless executor further coordinates transmission scheduling, subband assignment, and power allocation under heterogeneous stragglers. Trace-driven evaluations show that the proposed collective synthesis improves communication efficiency, while adaptive hybrid-fabric coordination achieves a more favorable completion-time and transmission energy tradeoff than baselines. 
% These results demonstrate the importance of jointly considering collective semantics, network state, and wireless resource constraints when integrating reconfigurable THz links into AI datacenter networks.

% \begin{figure}[!t]
%     \centering
%     \includegraphics[width=\columnwidth]{fig/a2a_straggler.png}
%     \caption{Completion time of A2A under compute stragglers.}
%     \label{fig:a2a_straggler}
% \end{figure}

% \begin{figure}[!t]
%     \centering
%     \includegraphics[width=\columnwidth]{fig/a2a_no_straggler.png}
%     \caption{Completion time of A2A without compute stragglers.}
%     \label{fig:a2a_no_straggler}
% \end{figure}

% \begin{figure}[!t]
%     \centering
%     \includegraphics[width=\columnwidth]{fig/allreduce_straggler.png}
%     \caption{Completion time of AllReduce under compute stragglers.}
%     \label{fig:allreduce_straggler}
% \end{figure}

% \begin{figure}[!t]
%     \centering
%     \includegraphics[width=\columnwidth]{fig/allreduce_no_straggler.png}
%     \caption{Completion time of AllReduce without compute stragglers.}
%     \label{fig:allreduce_no_straggler}
% \end{figure}

\clearpage

\bibliographystyle{IEEEtran}
\bibliography{ref}

\end{document}